\documentclass[aip,graphicx,preprint,onecolumn]{revtex4-2}

\usepackage {amssymb}
\usepackage {amsmath}
\usepackage{graphicx}

\draft 

\begin{document}


\title{Specific features of the $\pi$-electron spectrum of narrow achiral $(2m,m)$ nanoribbons} 



\author{Lyuba Malysheva}
\email[]{malysh@bitp.kyiv.ua}
\affiliation{Bogolyubov Institute for Theoretical Physics, Kyiv, Ukraine}


\date{\today}

\begin{abstract}

 On the basis of
the Su-Schrieffer-Heeger-H\"uckel-type Hamiltonian, we consider the tight-binding eigenvalue problem 
for a sequence of pyrene molecules forming a narrow $(2m,m)$ graphene nanoribbon. Specific features of the corresponding dispersion relation are analyzed and  illustrated with several examples.
It is shown
that the $\pi$-electron spectrum of the pyrene oligomer includes  local states, in contrast to  the spectrum of linear acene, which consists only of extended states. We analyze and illustrate the difference in the behavior of
 the electron density distribution for  extended and local electronic states.   Explicit analytic expressions for the Green's function coefficients of the pyrene molecule are also presented.
 \end{abstract} 


\maketitle 

\section{Introduction}
\begin{figure}
\centering
\includegraphics[width=0.79\textwidth]{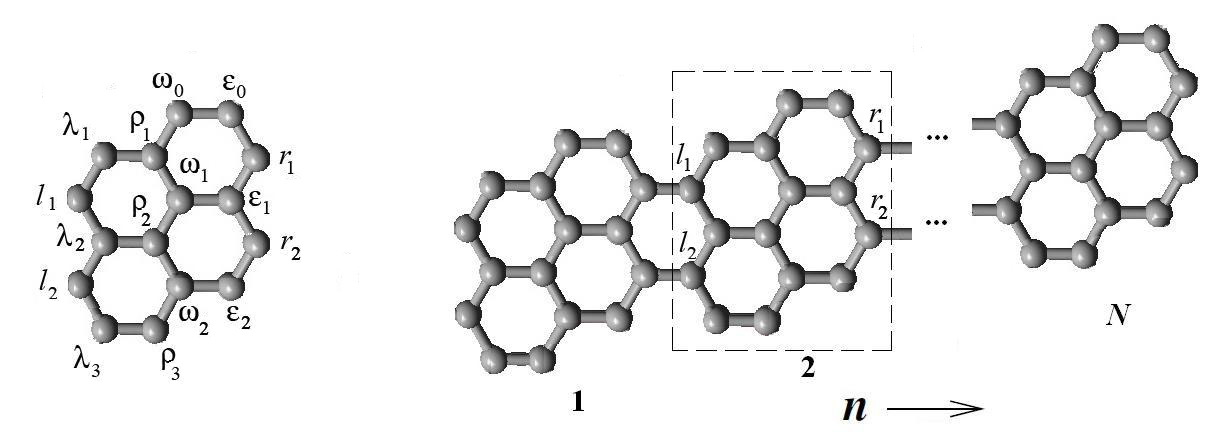}
\caption{  Labeling of the carbon atoms used in the proposed description of $\pi$-electron spectra of NGNR. The dash-framed block (pyrene molecule)  is the monomer forming $N-$length narrow $(2m,m)$ graphene nanoribbon. The elements of the Green's function matrix $g_{\alpha,\alpha'}$ of the monomer, presented in Appendix~\ref{A}, completely determine the electron spectrum of the NGNR.} \label{Fig1} 
\end{figure}

Electronic structure of narrow graphene nanoribbons  strongly depends on the conditions set at their boundaries. 
Along with a large number of studies devoted to unbounded graphene nanostructures 
 \cite{Wallace,Saito,Nakada,Fujita,Peres,Son,Geim,Saroka}, there are 
 numerous scientific works revealing various edge effects in finite-size graphene layers 
using different analytic and numerical methods \cite{Kobayashi,Klein1,Waka,Brey,Klein2,PRL,PRB,pss2013}, investigating, in particular,  the existence of edge (local) states in  graphene structures with  zigzag edges. Recently \cite{CPL}, the problem of existence of edge states in the nanostructures with nonarmchair and nonzigzag edges, i.e., in chiral nanoribbons, namely, in $(2m,m)$ graphene nanoribbons, was considered in the context of the electron structure of $\pi$-conjugated oligomers of polynaphthalene. 
In the present study we consider the specific features of $\pi$-electron spectra of narrow chiral $(2m,m)$ graphene nanoribbons, namely, of a sequence of pyrene molecules presented in  
Fig.~\ref{Fig1} and abbreviated below as NGNR. 

 Using the previously reported results \cite{PRL,PSS2008,CPL}, we consider a sequence of $N$ pyrene molecules (an isolated pyrene molecule is shown in the left panel of Fig.~\ref{Fig1}  and is dash-framed in the right panel), which are coupled with each other by two C--C covalent bonds. All dangling bonds along the nanoribbon edges are supposed to be filled by hydrogen atoms not shown in the figure and, thus, the considered model represents an ideal $\pi$-electron system. Thus, the electronic properties can be well described by the H\"uckel-type Hamiltonian with a single parameter, the C--C hopping integral between the nearest-neighbor carbon atoms, which is used in what follows as the energy unit, and the Fermi energy of $\pi$-electrons serves as the reference. 
 
 To  analyze the problem of existence of edge states in the bounded NGNR, we first write the general solution of the eigenvalue problem for corresponding  $\pi$-conjugated oligomers
 on the basis of
the Su-Schrieffer-Heeger-H\"uckel-type Hamiltonian,
  by using  the method of Lifshits \cite{Lifshits} and Koster and Slater \cite{Koster}.
We then apply the obtained relations to study the electronic properties of NGNR.  
 We obtain  exact analytic expressions for the coefficients of the  Green's function of the pyrene molecule, which are necessary for finding the dispersion relation of NGNR and present  the results of calculations of electronic spectra of a sequence of  pyrene molecules with a given  length parameter $N$.   Differences in the behavior of
 the electron density distribution for  extended and localized electronic states are also analyzed and illustrated.

\section{Exact Solution of the Eigenvalue Problem for a Sequence of $N$ Monomers}\label{Sec1}

By using the $n,\alpha$ labeling explained in Fig.~\ref{Fig1} (where $\alpha=l_j, r_j$, $j=1,2$; $\alpha=\lambda_j, \rho_j$, $j=1, 2, 3$, and $\alpha=\omega_j, \varepsilon_j$, $j=0,1,2$), the $\pi$-electron wave function in the Schr\"odinger equation $H\Psi=E\Psi$ with the tight-binding Hamiltonian $H$ of any oligomer formed by  $N$ similar monomers, can be represented in the following form:
\begin{equation}\label{1}
\Psi = \sum_{n=1}^{N}\sum_{\alpha }
\psi_{n,\alpha} |n,\alpha\rangle,
\end{equation}
where $|n,\alpha\rangle$ is a $2p_z$ orbital at the $\alpha$th atom of the $n$th monomer.
Substituting this expansion in the  Su-Schrieffer-Heeger-H\"uckel-type Hamiltonian of the oligomer presented  in Fig.~\ref{Fig1}
 and using  the method of Lifshits \cite{Lifshits} and Koster and Slater \cite{Koster}, we derive the following equation:
\begin{equation}\label{2}
\psi_{n,\alpha}=-\big (g_{\alpha,l_1}\psi_{n-1,r_1}+g_{\alpha,l_2}\psi_{n-1,r_2}\big)(1-\delta_{n,1}) -\big (g_{\alpha,r_1}\psi_{n+1,l_1}+g_{\alpha,r_2}\psi_{n+1,l_2} \big)(1-\delta_{n,N}),
\end{equation}
where the energy is expressed in the units of $\beta>0$ and 
$-\beta$ is the C--C hopping integral. Setting in Eq. (\ref{2}) $\alpha=l_1, r_1, l_2,$ and $r_2$,
we get a closed system of recurrence equations
for coefficients $\psi_{n,\alpha}$, $\alpha = l_1, r_1, l_2,$ and $r_2$:
\begin{align}
\psi_{n,l_1}&=-\big (g_{l_1,l_1}\psi_{n-1,r_1}+g_{l_1,l_2}\psi_{n-1,r_2}\big )(1-\delta_{n,1})-\big (g_{l_1,r_1}\psi_{n+1,l_1}+g_{l_1,r_2}\psi_{n+1,l_2} \big )(1-\delta_{n,N}), \notag \\
\psi_{n,l_2}&=-\big (g_{l_2,l_1}\psi_{n-1,r_1}+g_{l_2,l_2}\psi_{n-1,r_2}\big )(1-\delta_{n,1})-\big (g_{l_2,r_1}\psi_{n+1,l_1}+g_{l_2,r_2}\psi_{n+1,l_2} \big )(1-\delta_{n,N}), \notag \\[-15pt]\label{3}
\\[-15pt]
\psi_{n,r_1}&=-\big (g_{r_1,l_1}\psi_{n-1,r_1}+g_{r_1,l_2}\psi_{n-1,r_2}\big)(1-\delta_{n,1})-\big (g_{r_1,r_1}\psi_{n+1,l_1}+g_{r_1,r_2}\psi_{n+1,l_2} \big )(1-\delta_{n,N}), \notag \\
\psi_{n,r_2}&=-\big(g_{r_2,l_1}\psi_{n-1,r_1}+g_{r_2,l_2}\psi_{n-1,r_2}\big )(1-\delta_{n,1})-\big (g_{r_2,r_1}\psi_{n+1,l_1}+g_{r_2,r_2}\psi_{n+1,l_2} \big )(1-\delta_{n,N}). \notag 
\end{align}
We solve  system (\ref{3}) by introducing the generating functions
$$
L_{\left\{\substack{1 \\2}\right\}}=\sum_{n=1}^N z^n\psi_{n,l_{\left\{\substack{1 \\2}\right\}}}, \quad R_{\left\{\substack{1 \\2}\right\}}=\sum_{n=1}^N z^n\psi_{n,r_{\left\{\substack{1 \\2}\right\}}},
$$
satisfying the following equation:
\begin{equation}\label{4}
{\mathrm M}\begin{pmatrix}L_1\\L_2\\R_1\\R_2 \end{pmatrix}
=\begin{pmatrix}h_1\\h_2\\h_3\\h_4 \end{pmatrix},
\end{equation}
where
\begin{equation}\label{5}
{\mathrm M}=\begin{bmatrix}\Big(1+\dfrac{g_{l_1,r_1}}z\Big)&\dfrac{g_{l_1,r_2}}z&g_{l_1,l_1}z&g_{l_1,l_2}z \\
\dfrac{g_{l_2,r_1}}z&\Big(1+\dfrac{g_{l_1,r_1}}z\Big)&g_{l_1,l_2}z&g_{l_2,l_2}z\\
\dfrac{g_{l_2,r_2}}z&\dfrac{g_{l_1,l_2}}z&\Big(1+g_{l_1,r_1}z\Big)&g_{l_2,r_1}z\\
\dfrac{g_{l_1,l_2}}z&\dfrac{g_{l_1,l_1}}z&g_{l_1,r_2}z&\Big(1+g_{l_1,r_1}z \big)
 \end{bmatrix}
\end{equation}
and
\begin{align*}
h_1&=g_{l_1,r_1}\psi_{1,l_1}+g_{l_1,r_2}\psi_{1,l_2}+z^{N+1}g_{l_1,l_1}\psi_{N,r_1}+z^{N+1}g_{l_1,l_2}\psi_{N,r_2},\\
h_2&=g_{l_2,r_1}\psi_{1,l_1}+g_{l_1,r_1}\psi_{1,l_2}+z^{N+1}g_{l_1,l_2}\psi_{N,r_1}+z^{N+1}g_{l_2,l_2}\psi_{N,r_2},\\
h_3&=    g_{l_2,l_2}\psi_{1,l_1}+g_{l_1,l_2}\psi_{1,l_2}+z^{N+1}g_{l_1,r_1}\psi_{N,r_1}+z^{N+1}g_{l_2,r_1}\psi_{N,r_2},\\
h_4&=g_{l_1,l_2}\psi_{1,l_1}+g_{l_1,l_1}\psi_{1,l_2}+z^{N+1}g_{l_1,r_2}\psi_{N,r_1}+z^{N+1}g_{l_1,r_1}\psi_{N,r_2}.
\end{align*}
The dispersion relation for the considered oligomer is defined by the equation $\det({\mathrm M})=0$.
Thus, the energy-dependent quantities $g_{\alpha,\alpha'}$, where $\alpha, \alpha' = l_1, r_1, l_2, r_2,$ i.e., the  elements of the Green's function matrix referred to the same or different binding atoms of the monomer (presented in Appendix~\ref{A}) specify
the structure of the electron spectrum of narrow $(2m,m)$ carbon nanotubes, which is discussed in the next  section.

\begin{figure}
\centering
\includegraphics[width=0.6\textwidth]{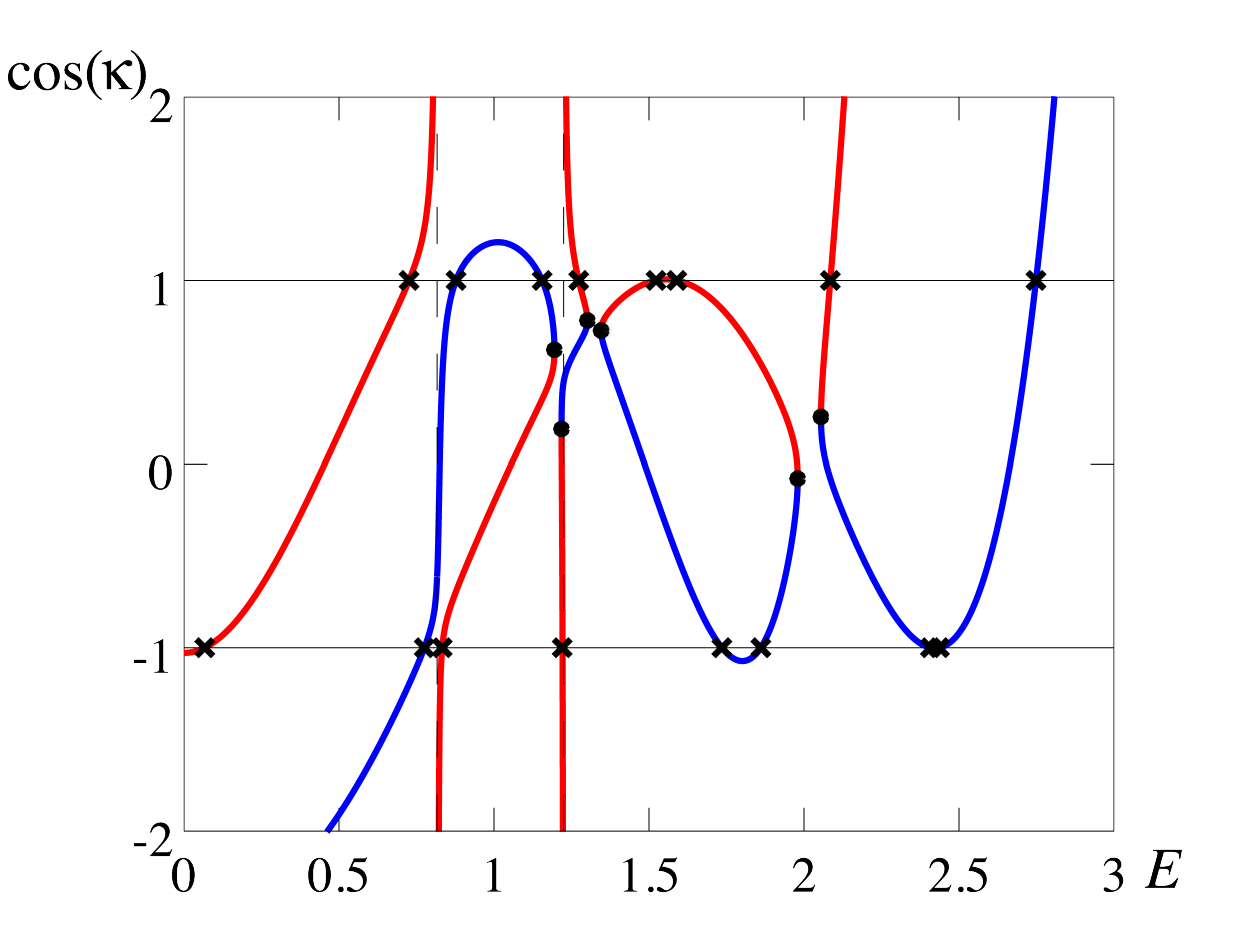}
\caption{ Dependences of $\cos\kappa_1$ (red curves) and $\cos\kappa_2$ (blue curves) on energy  (in units of the resonance integral)  calculated by using Eq. (\ref{6}) and the explicit relations for the  elements of the Green's function matrix
$g_{\alpha,\alpha'}$ derived in Appendix~\ref{A}. 
The dashed vertical lines  mark the energies  
$E^2=2/3$  and $E^2=3/2$, see text. 
 The band-edge energies  corresponding to
  $\cos \kappa_{ \left\{ \begin{smallmatrix} 1\\2 \end{smallmatrix} \right\}} =\pm1$ 
are marked by
crosses. Black dots correspond to the energies, for which $b^2=4ac$, i.e., $\cos\kappa_1=\cos\kappa_2$. }\label{Fig2}
\end{figure}

\section{Electron spectrum of a sequence of pyrene molecules}\label{Sec2}
\begin{figure}[t!]
\centering
\includegraphics[width=0.55\textwidth]{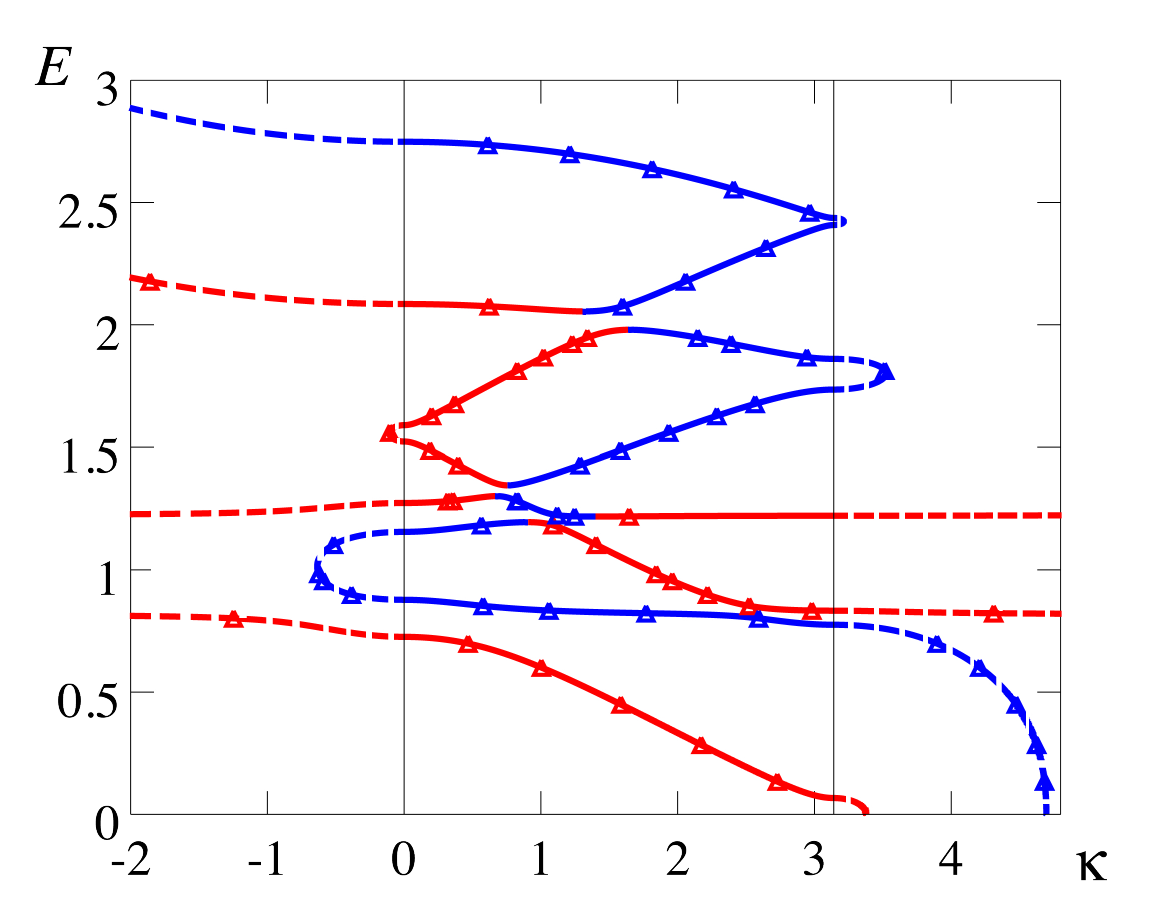}
\caption{Two branches of dispersion relations $E(\kappa_1)$ (red curves) and $E(\kappa_2)$ (blue curves) are presented only for positive energies (the curves $E(\kappa)$ are symmetric about the sign of $E$). The red  and blue triangles mark the eigenvalues for a bounded sequence of pyrene molecules with $N=5$. Schematic continuations to the imaginary 
values of wave vectors, $\kappa\rightarrow -i\delta$ and $\kappa\rightarrow \pi+i\delta$, $\delta\geq 0$,  are shown by dashed lines. The values $\kappa=0$ and $\kappa=\pi$ are shown by the vertical lines.}\label{Fig3}
\end{figure}

Denoting
$$
z+\dfrac1z=2\cos\kappa,
$$
we can write equation $\det({\mathrm M})=0$ in the form
\begin{equation}\label{6}
\big ( 2\cos\kappa \big )^2 g^{nd}_\Delta + 4\cos\kappa \big ( g_{l_1,r_1}
Q_1-g_{l_1,l_2}Q_2  \big ) 
+(Q_1 )^2- (Q_2 )^2-4\big(g_{l_1,l_2}^2-g_{l_1,r_2}g_{l_2,r_1}\big )
=0,
\end{equation}
where
$$
Q_1\equiv 1+g^{nd}_\Delta+g^d_\Delta, \quad Q_2\equiv 2g_{l_1,r_1}g_{l_1,l_2}-g_{l_2,r_1}g_{l_1,l_1}-g_{l_1,r_2}g_{l_2,l_2}, \quad g_\Delta^{d}=g_{l_1,l_2}^2-g_{l_1,l_1}g_{l_2,l_2},
$$
and
$$
g_\Delta^{nd}=g_{l_1,r_1}^2-g_{l_1,r_2}g_{l_2,r_1}=\dfrac{6E^4-13E^2+6}{(E^2-4)(E^2-1)(E^6-5E^4+6E^2-1)(E^6-9E^4+18E^2-9)}.
$$
Equation (\ref{6}) is obviously a quadratic equation 
$$a\big ( 2\cos\kappa \big )^2+b\big ( 2\cos\kappa \big )+c=0$$
 with respect to $\cos\kappa$ and evidently has two solutions 
\begin{equation}\label{7}
\cos\kappa_{\left\{\substack{1 \\2}\right\}}=\dfrac{-b/2\pm\sqrt{b^2/4-ac}}a.
\end{equation}
  These two solutions determine two branches of dispersion relation $E(\kappa)$ and are presented in  Figs.~\ref{Fig2} and ~\ref{Fig3}. It is instructive to compare these dependences with the  two branches of dispersion relation for linear acene: $4\cos^2\big(\kappa_{\left\{\substack{1 \\2}\right\}}/2\big)=E(E\pm1)$ (see, e.g., Ref. \cite{PSS2008}).

\section{Specific features of the electron structure}
 By analyzing Eq. (\ref{6}) and the explicit relations for  $g_{\alpha, \alpha'}$ we derive the following properties of the  dependences $E(\kappa)$:

\begin{enumerate}
\item[ (i)]
The NGNR spectrum is symmetric about the sign of $E$.

\item[(ii)]
For $E=0$, it is easy to obtain
$$
 2\cos\kappa_{\left\{\substack{1 \\2}\right\}} =\dfrac{-21\pm5\sqrt{3}}{6}=\left \{ \begin{matrix} -2.057\\-4.943 \end{matrix}\right., \quad  \kappa_{\left\{\substack{1 \\2}\right\}}=\left \{ \begin{matrix}\pi -0.237i\\\pi-1.554i.\end{matrix}\right.
$$

\item[ (iii)]
For $E=\pm 1$, 
we get 
$$
2\cos\kappa_{\left\{\substack{1 \\2}\right\}} =1\mp\sqrt{2}=\left \{ \begin{matrix} -0.414\\ 2.414 \end{matrix}\right., \quad  \kappa_{\left\{\substack{1 \\2}\right\}}=\left \{ \begin{matrix} 1.779 \\ 0.633i\end{matrix}\right. .
$$

\item[(iv)]
If $a=0$, i.e., for  $E^2=2/3$ and $E^2=3/2$, we get
$
2\cos\kappa_{1}=2\cos\kappa_2=-c/b,
$ 
or
$
 2\cos\kappa_{1}=2\cos\kappa_2=-1.226, \quad \kappa_1=\kappa_2=2.231
$  for $E^2=2/3$
and 
$
2\cos\kappa_{1}=2\cos\kappa_2=0.896, \quad \kappa_{1,2}=1.106
$
for $E^2=3/2$.

\item[(v)]
$\cos \kappa=1$ for 
$E=0.7260, 0.8775, 1.155, 1.273, 1.523, 1.590, 2.085$, and 2.748; these values are marked by crosses in Fig.~\ref{Fig2}. 

\item[ (vi)]
$\cos \kappa=-1$ for 
$E=0.067, 0.775, 0.833, 1.221, 1.735, 1.861, 2.408$, and 2.437, as is shown in  Fig.~\ref{Fig2}. 

\end{enumerate}
It is clear that the energy regions, where $|\cos \kappa| > 1$, do not contain extended eigenvalues.

Our calculations performed  for the solution of system (\ref{4}) show that the NGNR spectrum  
contains both extended and localized energy levels. For the particular case of a sequence of 5 pyrene molecules, this is illustrated in Fig.~\ref{Fig3}. 
Using these results, we analyze the   behavior of
 the electron density distribution for  extended and localized electron states in the next section.

\section{Coefficients of expansion of the wave functions}
By analyzing the  expansion coefficients of the $\pi$-electron wave function (\ref{1}) satisfying Eq.(\ref{2}), we can determine  the behavior  of the electron density distribution and trace the   difference between the probabilities of finding an electron in a certain position for  extended and local energy eigenvalues.

\begin{figure}[t!]
\centering
\includegraphics[width=0.9\textwidth]{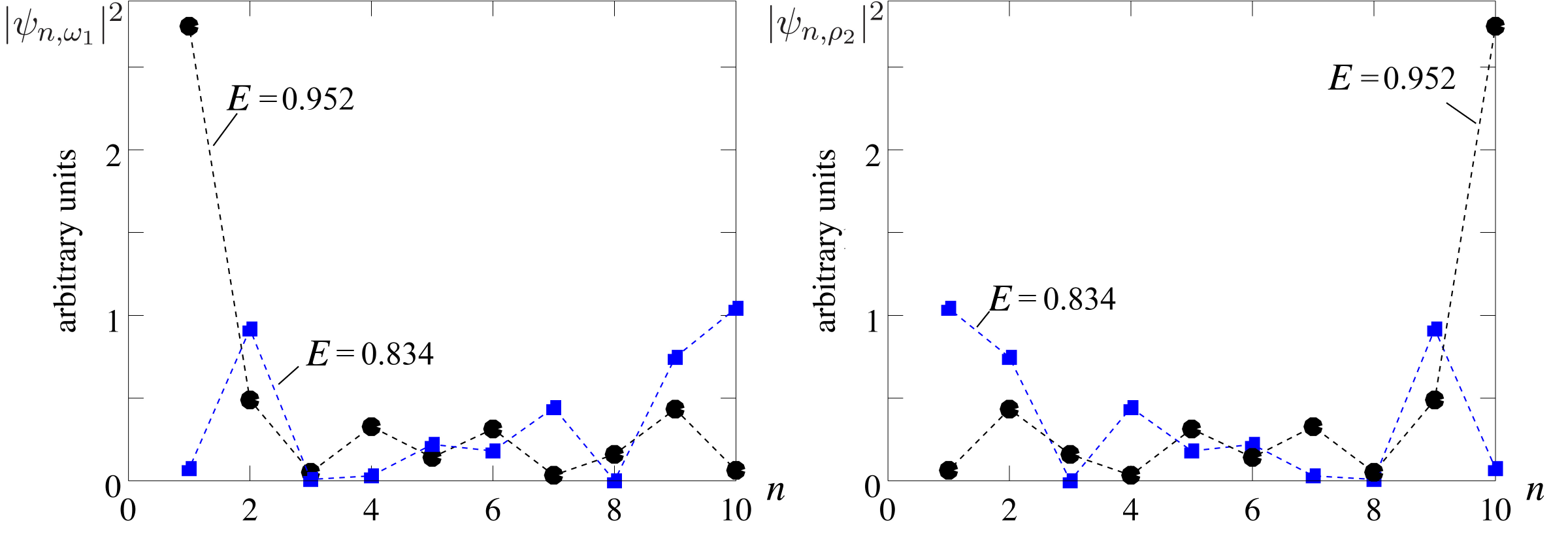}
\caption{Squared modulus of expansion coefficients of the wave
functions $|\psi_{n,\omega_1}|^2$ (left panel) and $|\psi_{n,\rho_2}|^2$ (right panel) as a function of $n$ for $N = 10$. The black circles correspond to the eigenvalue $E=0.952$ ($\kappa_1=1.972$, $\kappa_2= - 0.583i$), while the blue squares correspond to the eigenvalue $E=0.834$ ($\kappa_1=2.943$, $\kappa_2= 1.043$).}\label{Fig4}
\end{figure}

In Fig.~\ref{Fig4}, we present the squared modulus  $|\psi_{n,\omega_1}|^2$ (left panel) and $|\psi_{n,\rho_2}|^2$ (right panel) for the following eigenvalues: $E=0.952$ (extended state $\kappa_1=1.972$ and localized state $\kappa_2= - 0.583i$)  and $E=0.834$ (extended states $\kappa_1=2.943$, $\kappa_2= 1.043$). The   squared wavefunction coefficients for the extended eigenvalue demonstrate an oscillating dependence on $n$. In contrast, a rather sharp increase in the square of the modulus for the localized state with increasing $n$ is observed.
 Thus, for  localized states,
 the calculated electron density distribution demonstrates the presence of rather sharp maxima at the ends of a bounded sequence of pyrene molecules.  A similar  difference in the behavior of
 the electron density distribution for the extended and local electron states was recently reported for bounded polynaphthalene oligomers and can be considered
as a specific feature of narrow chiral $(2m,m)$ graphene nanoribbons.

\appendix

\section{ Green's functions for the pyrene molecule}\label{A}

The Green's functions $g_{\alpha, \alpha'}$ were found  by using the  Symbolic Math Toolbox in the MATLAB software package:
$$
D=(E^2-4)(E^2-1)(E^6-5E^4+6E^2-1)(E^6-9E^4+18E^2-9), 
$$
$$
g_{l_1,r_1}D=g_{l_2,r_2}D=-(E^2-1)(3E^8-23E^6+58E^4-57E^2+18),
$$
$$
g_{l_1,l_1}D=g_{r_2,r_2}D=E(E^{14}-17E^{12}+112E^{10}-370E^8+660E^6-631E^4+295E^2-51),
$$
$$
g_{l_1,l_2}D=E(E^2-1)(E^{10}-13E^8+61E^6-128E^4+120E^2-39),
$$
$$
g_{l_2,l_2}D=g_{r_1,r_1}D=E(E^2-1)(E^{12}-16E^{10}+96E^8-275E^6+392E^4-261E^2+63)
$$
$$
=E(E^2-1)^2(E^2-3)(E^8-12E^6+45E^4-59E^2+21),
$$
$$
g_{l_1,r_2}D=-(3E^{10}-23E^8+56E^6-49E^4+6E^2+6),
$$
$$
g_{l_2,r_1}D=-(E^2-1)(E^8-E^6-18E^4+39E^2-18).
$$

The following combinations are used in the main text:
$$
g_\Delta^{nd}=g_{l_1,r_1}^2-g_{l_1,r_2}g_{l_2,r_1}=\dfrac{6E^4-13E^2+6}{D},
$$
$$
g_\Delta^{d}=g_{l_1,l_2}^2-g_{l_1,l_1}g_{l_2,l_2}=-E^2\dfrac{E^{12} -15E^{10}+84E^8-225E^6+304E^4-196E^2+47}{D}.
$$

\end{document}